\documentclass[useAMS,usenatbib]{mn2e}
\usepackage{graphicx}

\begin{document}

\def \th {\thinspace}
\def \arcmin {\hbox{$^\prime$}}
\def \arcsec {\hbox{$^{\prime\prime}$}}
\def \chisq {$\chi ^{2}$}
\def\approxgt{\mathrel{\hbox{\rlap{\lower.55ex \hbox {$\sim$}} \kern-.3em \raise.4ex \hbox{$>$}}}}
\def\lesssim{\mathrel{\hbox{\rlap{\lower.55ex \hbox {$\sim$}} \kern-.3em \raise.4ex \hbox{$<$}}}}
\def\approxlt{\mathrel{\hbox{\rlap{\lower.55ex \hbox {$\sim$}} \kern-.3em \raise.4ex \hbox{$<$}}}}
\def \degmark {^\circ}
\def \sun {\hbox {$\odot$}}

\title[Gradient and dispersion analyses of the WMAP data]
{Gradient and dispersion analyses of the WMAP data}

\author[K.T. Chy\.zy, B. Novosyadlyj, M. Ostrowski]
{K.T. Chy\.zy$^1$\thanks{E-mail: K.Ch.: chris@oa.uj.edu.pl ; M.O.:  mio@oa.uj.edu.pl}, 
B. Novosyadlyj$^2$\thanks{E-mail: novos@astro.franko.lviv.ua}, M. Ostrowski$^1$
 \\
$^1$Astronomical Observatory of the Jagiellonian University, ul. Orla 171, 30-244 Krak\'{o}w, Poland\\
$^2$Astronomical Observatory of the Ivan Franko National University of Lviv, Kyryla i Methodia str., 8, Lviv, 79005, Ukraine}  

\date{  }


\maketitle

\label{firstpage}

\begin{abstract}
We studied the WMAP temperature anisotropy data using two different methods. The 
derived signal gradient maps show regions with low mean gradients in structures 
near the ecliptic poles and higher gradient values in the wide ecliptic equatorial 
zone, being the result of non-uniform observational time sky coverage. We show 
that the distinct observational time pattern present in the raw (cleaned) 
data leaves also its imprints on the composite CMB maps. Next, studying  
distribution of the signal dispersion we show that the north-south asymmetry of 
the WMAP signal diminishes with galactic altitude, confirming the earlier 
conclusions that it possibly reveals galactic foreground effects. As based on 
these results, one can suspect that the  instrumental noise sky distribution and 
non-removed foregrounds can have affected some of the analyses of the CMB signal. 
We show that actually the different characteristic axes of the CMB sky distribution 
derived by numerous authors are preferentially oriented towards some distinguished 
regions on the sky, defined by the observational time pattern and the galactic 
plane orientation. 
\end{abstract}


\section{Introduction}
The first-year data from the Wilkinson Microwave Anisotropy Probe (WMAP) 
(Bennett at al. 2003a,b) became a basis for testing cosmological 
scenarios, substantial improving of evaluation accuracy of basic cosmological 
parameters related to the geometry of our Universe, dynamics of its expansion, matter 
and energy contents. Therefore any studies of reliability of the 
provided data are of great interest and importance. In particular, application 
of non-standard methods provide additional independent tests of data 
quality and consistency, as well as the correctness of involved interpretations. 

All the sky CMB anisotropy maps give possibility to test with high accuracy the 
fundamental principles of current cosmology, such as the isotropy of our Universe 
and the Gaussianity of primordial cosmological perturbations. The first analysis of 
WMAP data for Gaussianity was performed by WMAP team 
\cite{komatsu2003}. They found that the WMAP data are consistent with the Gaussian 
primordial fluctuations and provided limits to the amplitude of 
non-Gaussian ones. Other authors, however,  applying advanced or sophisticated 
statistical methods to analyse the foreground-subtracted WMAP maps pointed 
out  non-Gaussian signatures or north-south asymmetry of galactic hemispheres  
(Chiang et al. 2003, Park 2004, Eriksen et al. 2004a, Hansen et al. 2004a,b,c, 
Eriksen et al. 2005a, Cayon et al. 2005). Some authors suppose that it may 
be caused by residual foreground contamination from unknown galactic 
or extra-galactic  sources and find their possible locations, angular 
scales, and amplitudes (Park 2004, Eriksen et al. 2004a, Hansen et al. 2004b, Chiang
\& Naselsky 2004, Patanchon et al. 2004, Naselsky et al. 2005, Tojeiro et al. 2005, 
Land \& Magueijo 2005). For example, Tojeiro et 
al. (2005) argued that the evidence for non-Gaussianity on large scales is 
associated with cold spots of unsubtracted  foregrounds. Some part of 
non-Gaussian  statistics of $\Delta T/T$ fluctuations can be due to the ring of 
pixels of radius of 5 degrees centered at ($l\approx 209^o, b\approx -57^o$) 
\cite{cayon2005}, close to the spot detected in Vielva et al. (2004) and Cruz et 
al. (2005). Wibig \& Wolfendale (2005) performed the correlation analysis of 
CMB temperature and gamma-ray whole-sky maps and detected positive 
correlation, which may suggest an actual contamination of the WMAP data by, if not 
cosmic rays directly, some component accompanying them in interstellar 
space with a sufficiently flat emission spectrum. 

All these studies indicate that WMAP Q-, V-, and W-band maps contain information 
on unidentified extended low-amplitude foreground structures that contaminate 
the relic CMB signal. The effective extraction of these structures from the WMAP 
maps would improve the current and future CMB analyses, as well as reveal possible 
new foreground objects or effects (cf. Davies et al. 2005, Eriksen et al. 2005b).    

Our principal goal in the present paper is to understand the physical meaning of 
structures within WMAP measurements in order to detect or limit possible 
non-cosmological factors affecting the data used for determining cosmological 
parameters. We apply two simple analytical methods to study possible foreground 
or observational effects in the WMAP first-year data, not recognized and/or p
removed by the cleaning methods used by the WMAP team (Bennett et al. 2003b). In 
the paper, first we shortly provide the basic information about the data analysed, 
and next describe our method of generating gradient maps from the foreground-cleaned 
WMAP maps and interpret the large-scale gradient structures by the
observational strategy of the WMAP satellite. We demonstrate that the scanning effects 
present in the raw signal data can leave clear imprints on the gradients maps 
from all frequency channels, retained even in coadded, smoothed, and ILC maps. 
Then, we 
investigate distribution of signal dispersion in the WMAP maps to find 
a significant asymmetry between the northern and southern galactic hemispheres. 
Following some previous authors, we argue that a natural explanation of this 
asymmetry is residual galactic contamination. Finally, the inevitable influence 
of the above-mentioned effects on the analysis of the large-scale CMB structure 
is shortly considered.

\begin{figure*}                                                 
\includegraphics[width=55mm,height=84mm,angle=270]{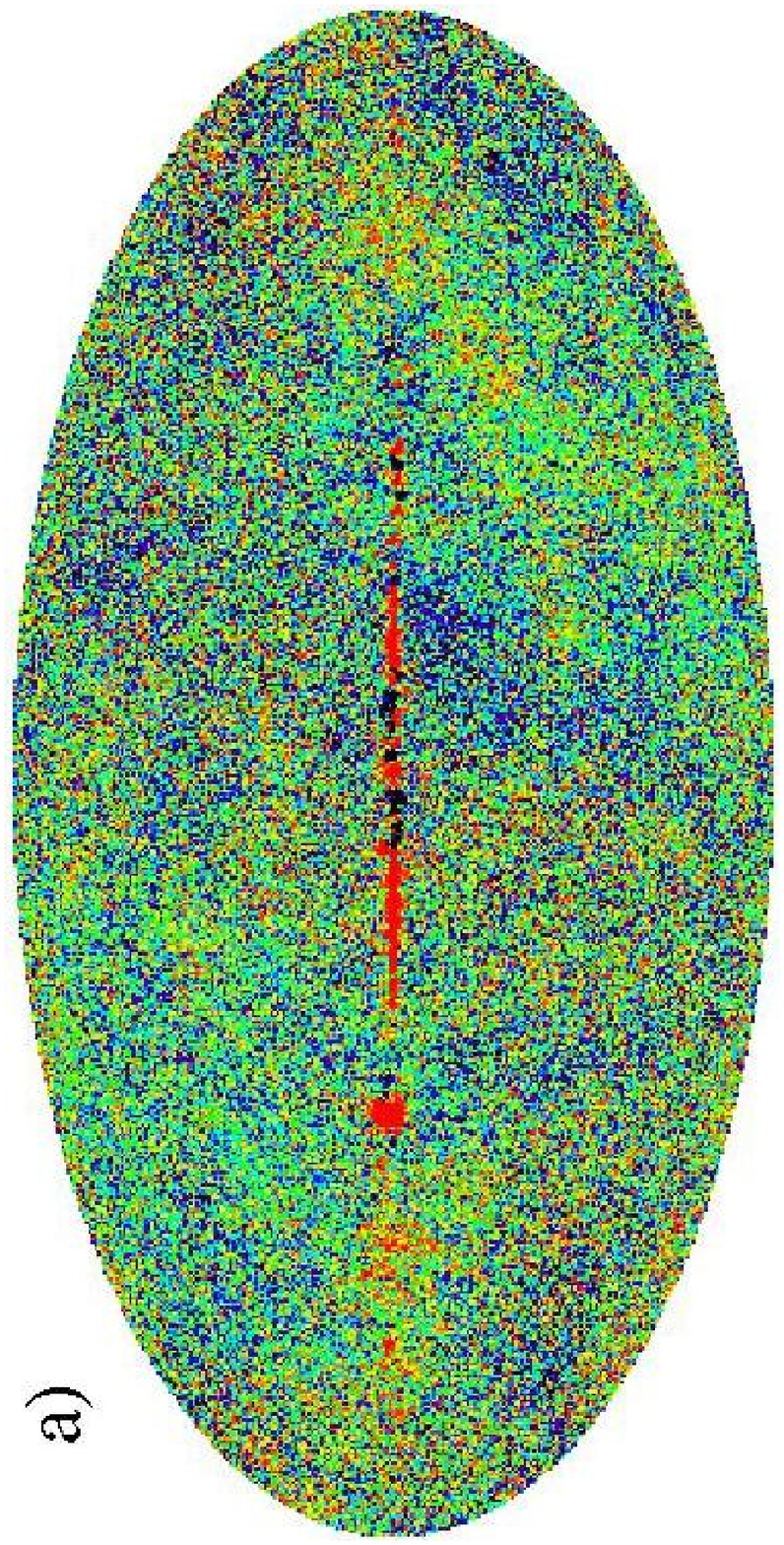}
\includegraphics[width=55mm,height=84mm,angle=270]{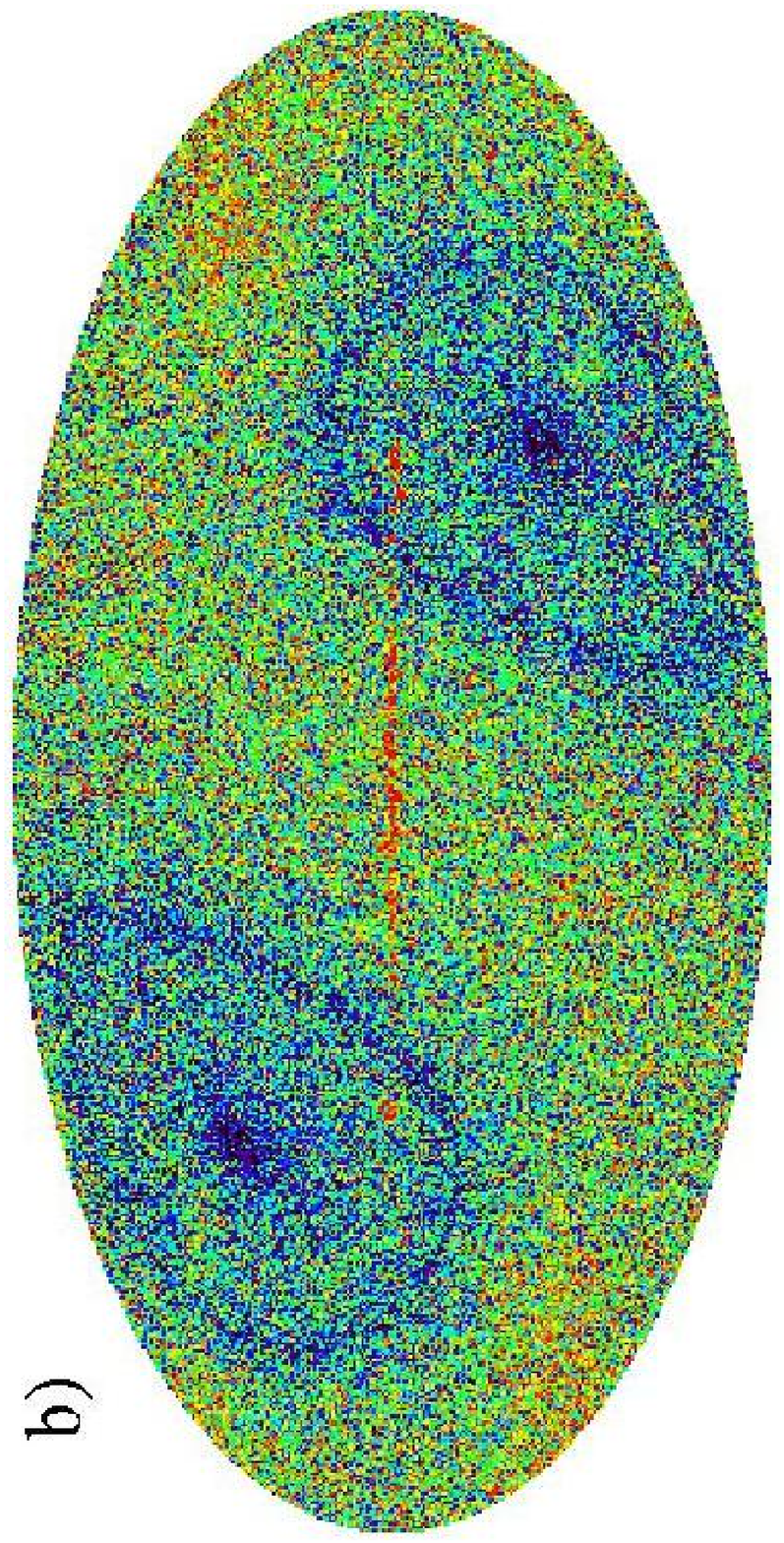}
\includegraphics[width=55mm,height=84mm,angle=270]{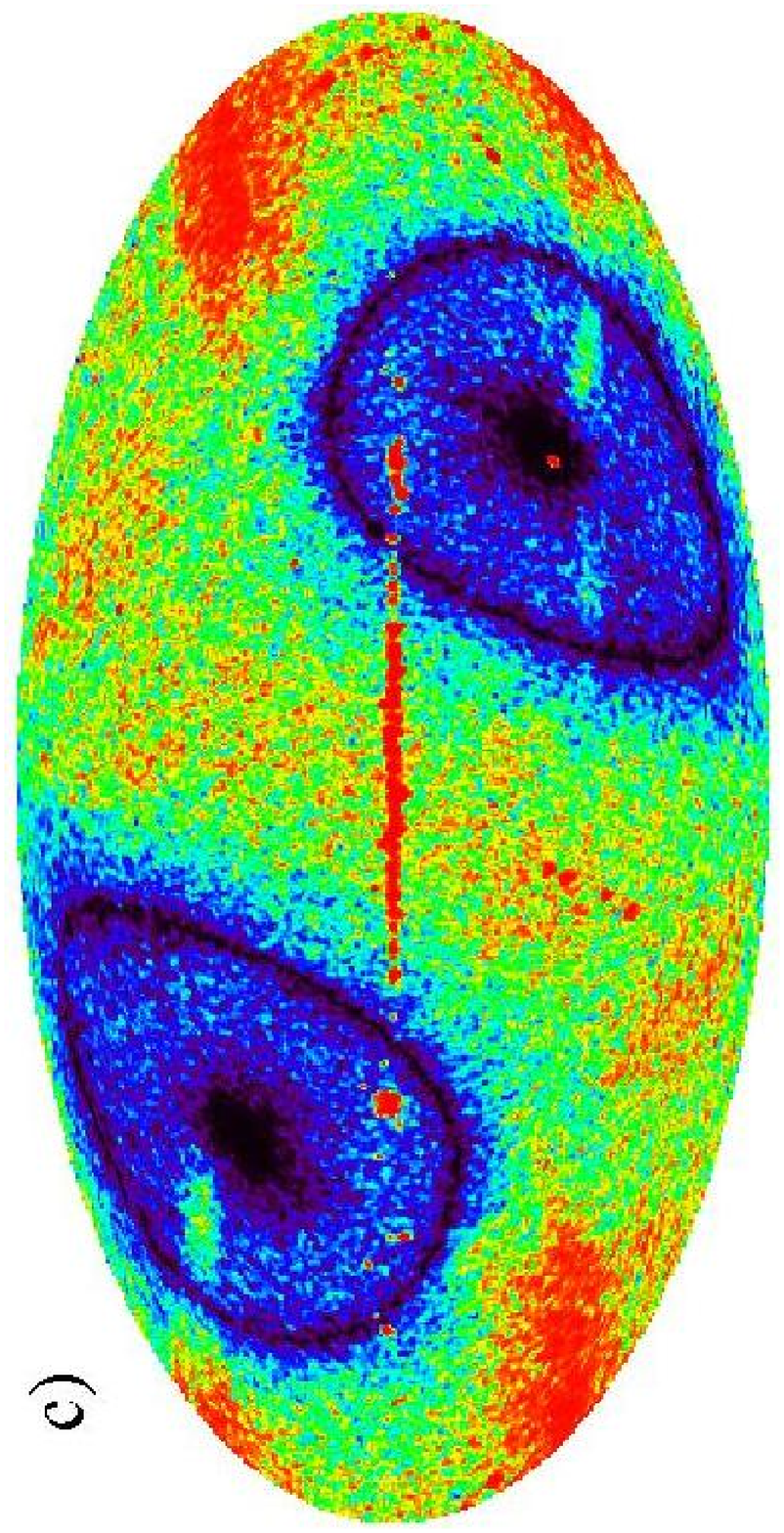}
\includegraphics[width=55mm,height=84mm,angle=270]{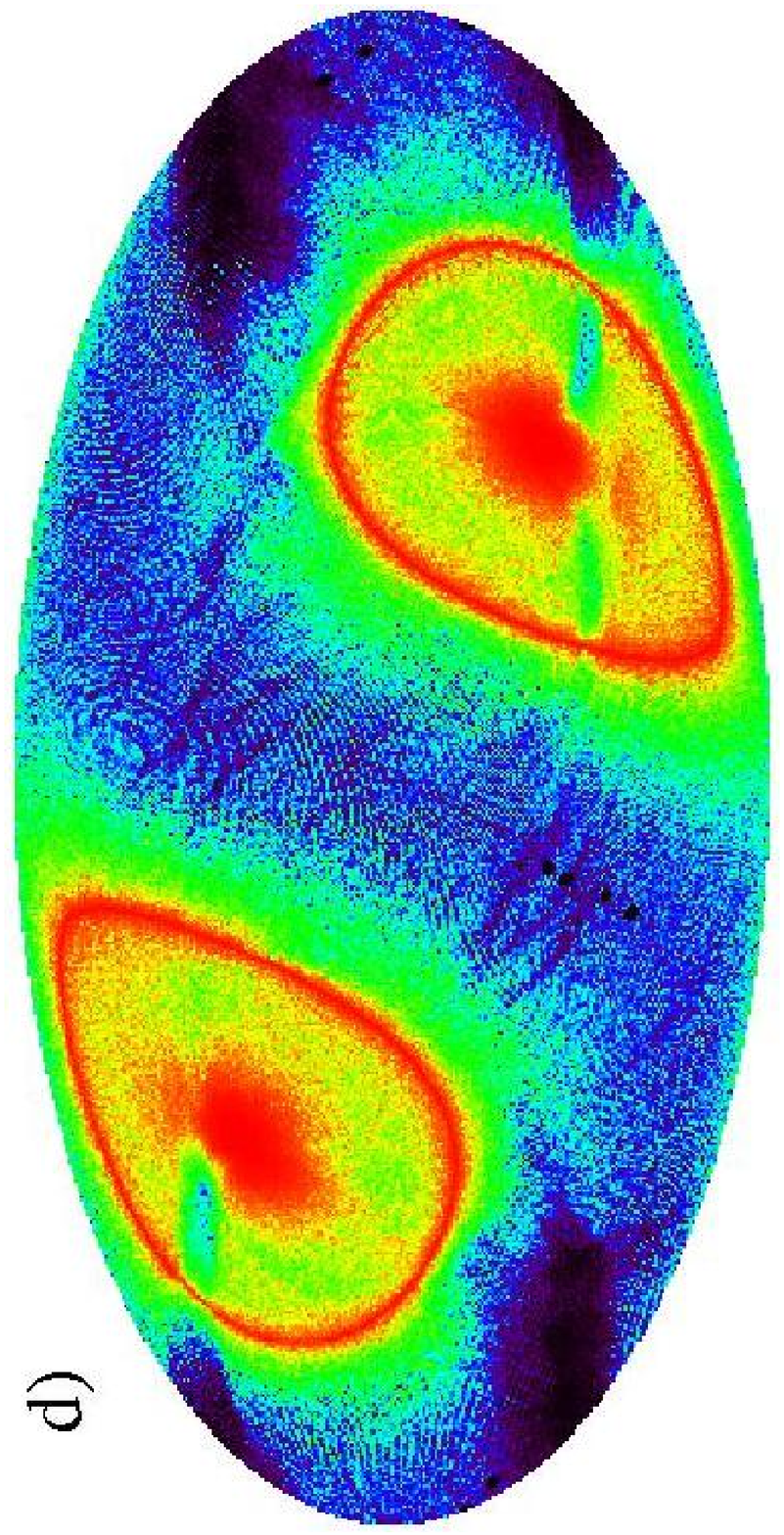}
\caption{
The maps in the Mollveide projection of the galactic coordinate
frame present: a) the signal $\Delta T$ in the WMAP V2 channel; b) the V2
gradient map derived from (a), blue color denotes low gradient values and red 
color - higher ones; c) the V2 gradient map smoothed out using the
$1\degr\,$ Gaussian beam; d) the effective number of observation map for the
channel V2; blue color denotes small numbers and red color - larger ones. 
All maps are presented in the histogram-scale format to enhance
the apparent features. The structures in the galactic plane have roughly about
ten times higher gradients than those outside it.
}
\label{fig1}
\end{figure*}

\section{The observational data}

In this paper we study the publicly available first-year WMAP maps in 5 
frequency channels, K (22.8 GHz), Ka (33.0 GHz), Q (40.7 GHz), V (60.8 GHz) 
and W (93.5 GHz) downloaded from LAMBDA\footnote{\tt http://www.lambda.gsfc.nasa.gov} 
website. The maps provide distributions of differential temperature measured 
in micro-Kelvin units, $\Delta T$, over the entire sky. The signal in each 
individual map, after subtracting the kinematic dipole component, foreground 
point sources, and diffuse galactic emission, is expected to consist primarily 
of cosmic microwave background (see Bennett et al. 2003a,b for details). 
We studied mainly 
individual differential assemblies maps with the full available resolution 
(NSIDE=512 in the HEALPix\footnote{\tt http://www.eso.org/science/healpix/} 
pixelization). Most analyses presented in the paper use the V2 channel map, 
showing the lowest galactic contamination and small instrumental noise.

\section{Gradient analysis of the WMAP CMB anisotropy maps}

\subsection{Derivation of gradient maps}

We study the variations of measured temperature fluctuations in 
the  WMAP sky maps, including CMB signal and instrumental noise, 
by analysing the gradient structure of the data. We applied two methods in 
order to derive the gradients. In the first approach, we evaluate the {\it maximum} 
gradient value at a given location in the map (i.e. in each pixel/cell) 
calculated between the reference cell and the surrounding cells using:

$$ ({\rm grad}\, \Delta T)_{0,max} = Max \left| {\Delta T_i - \Delta T_0 
\over d_{i,0}} \right| \qquad , \eqno(3.1)$$

\noindent
where the index "$0$" indicates the reference cell. "$i$" runs through 
all adjacent cells and $d_{i,0}$ is the distance between the cells considered . 
The second approach involved deriving a gradient map with the use of the Sobol 
operator (filter) known in the image analysis as the edge detector (Huang et al. 
2003). This method evaluated the {\it mean} gradient at a given pixel using 
the relative values of the signal, $\Delta T_i$, in its 8 adjacent pixels. In 
this case, the obtained gradient maps reveal the same pattern as in the maximum 
gradient method, but the derived mean gradient values are lower. Below we present 
the maximum gradient maps derived with the simpler method using eq. (3.1), 
which is more efficient in visualizing gradient structures. 

The map of $({\rm grad}\, \Delta T)_{0,max}$ derived for the WMAP V2 channel is 
presented in Fig.~1b. In the figure one can clearly see regions of low gradients at 
both the ecliptic poles and within two extended circular regions around the poles 
at the radius $\approx 23^\circ$. The extended regions of enhanced gradient occupy 
a wide stripe along the ecliptic equator. The largest gradients occur at the galactic 
plane due to some residual galactic component. Relatively large gradient values 
are also found in places where data were partly discarded because of the  
planets appearing in the WMAP field of view. The comparison of the gradient 
pattern with the map of observational time ($\equiv$ the {\it effective 
number of observations} in the WMAP team nomenclature, $N$) (Fig.1d) proves 
that most of the observed extended gradient structures represent variations 
of instrumental noise within the data. The noise distribution corresponds to the 
varying observational time in different areas of the sky due to the WMAP sky 
scanning pattern. It also leads to a distinct anti-correlation between the 
observational time and the signal gradient.
\begin{figure*}                                                 
\includegraphics[width=55mm,height=84mm,angle=270]{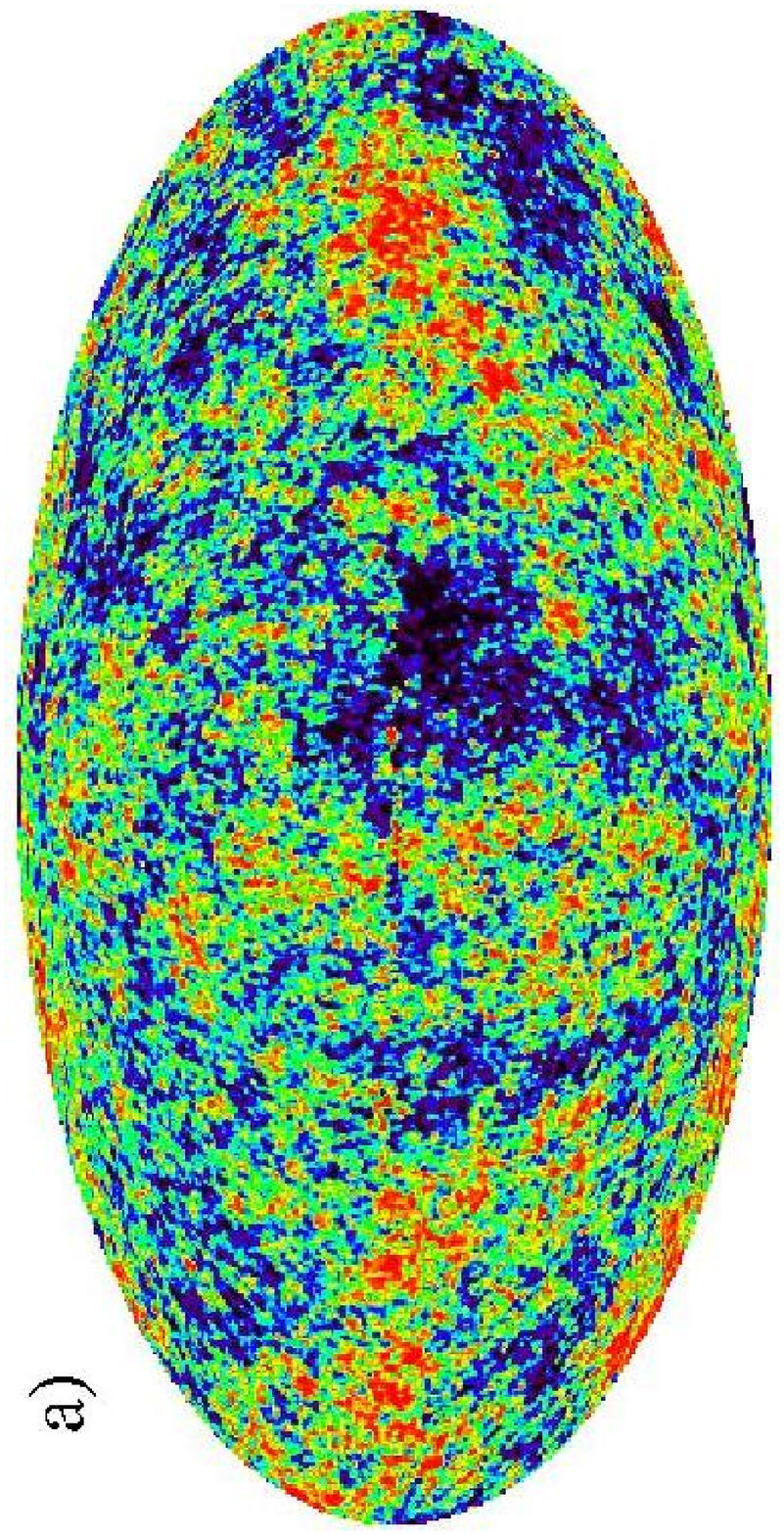}
\includegraphics[width=55mm,height=84mm,angle=270]{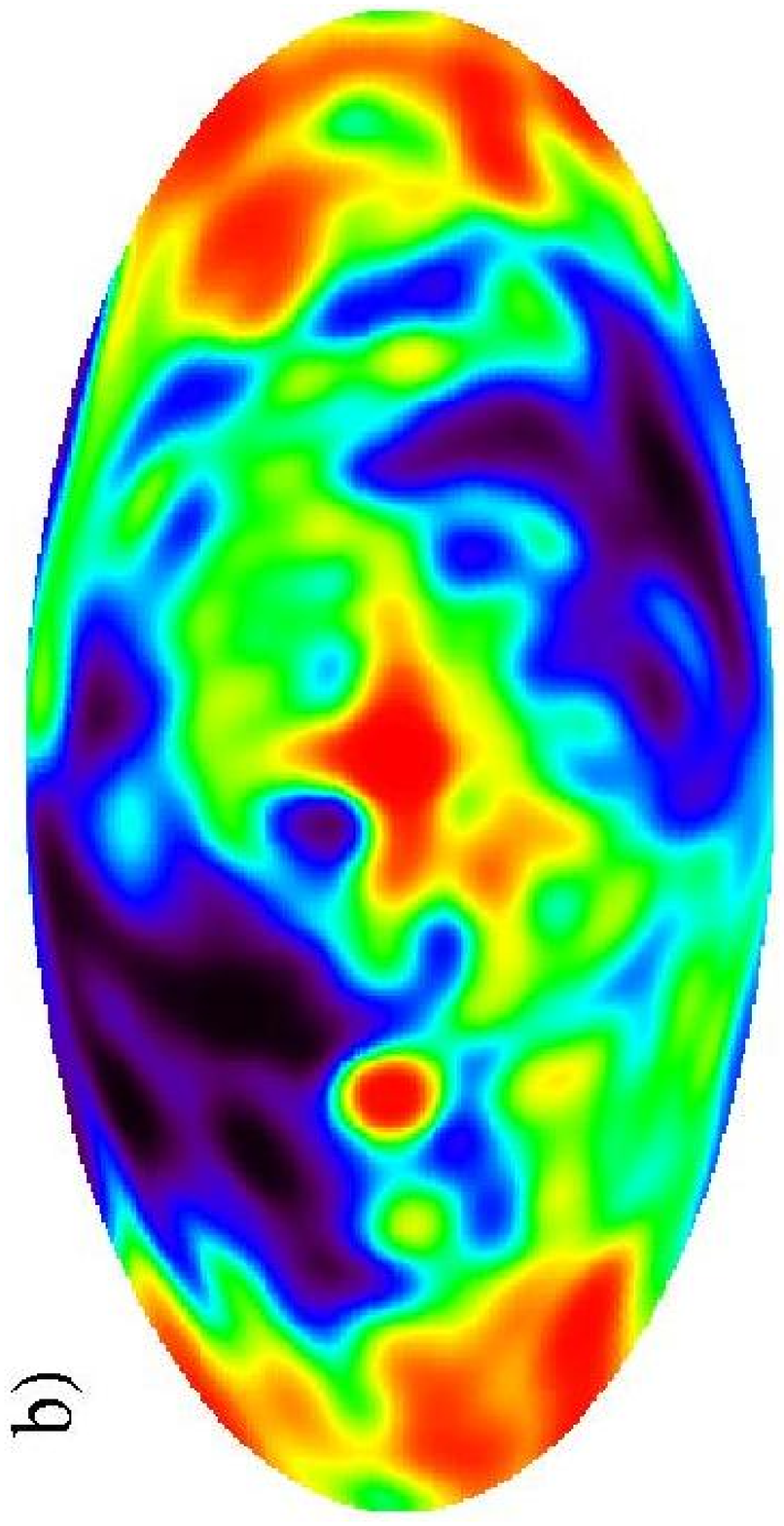}
\caption{ 
a) An ILC map of the CMB temperature fluctuations and  b) its
gradient map smoothed with the $15\degr\,$ Gaussian beam.
}
\label{fig2}
\end{figure*}

\subsection{Analysis of gradient pattern}

Using the above method, we derived the gradient maps for all the WMAP channels (the 
maps can be obtained upon request from K.Ch.) with analogous characteristic 
gradient patterns, spots at galactic poles, and rings around them. 
These characteristic features in the gradient distribution are more distinct 
after smoothing the maps (Fig.~1c). In order to show it quantitatively, we smoothed 
both the V2 CMB and the number of observation maps with increasing Gaussian 
beams, and calculated the correlation coefficient. Without smoothing, the 
correlation is -0.16, for 1\degr\ smoothing it is -0.26, and for 5\degr, 
10\degr, and 15\degr\ smoothing it becomes -0.69, -0.83, and -0.89, respectively.
 
We also obtained the gradient maps from the coadded maps for individual 
assemblies in V, Q and W bands, with the position-dependent, or band-dependent 
inverse noise-weighting. We got gradient patterns  much similar to those 
obtained from the V2 channel alone. Next we applied the gradient analysis to 
the ILC map. In this case, the characteristic gradient pattern is weaker, 
because of the convolved to 1\degr\ scale component maps of the ILC combined 
map. However, the distinct pattern appears also in this case, as we demonstrate 
at Fig.~2b by convolving the gradient map with the 15\degr\ Gaussian beam. In 
the figure, the observational time pattern analogous to that in Fig.~1d is clearly 
reproduced, while significantly perturbed by the effects of CMB or galactic 
foreground signal structures. 

The performed experiments confirm that the systematic pattern revealed in the 
gradient maps is firmly established and cannot be removed by simple filtering 
procedures, like convolving, or adding data maps. Thus it can affect the 
analysis and interpretation of the large-scale distribution of the signal 
in the WMAP maps.

\begin{figure*}                                                 
\begin{minipage}[b]{1\textwidth}
\includegraphics[width=6.8cm,angle=270]{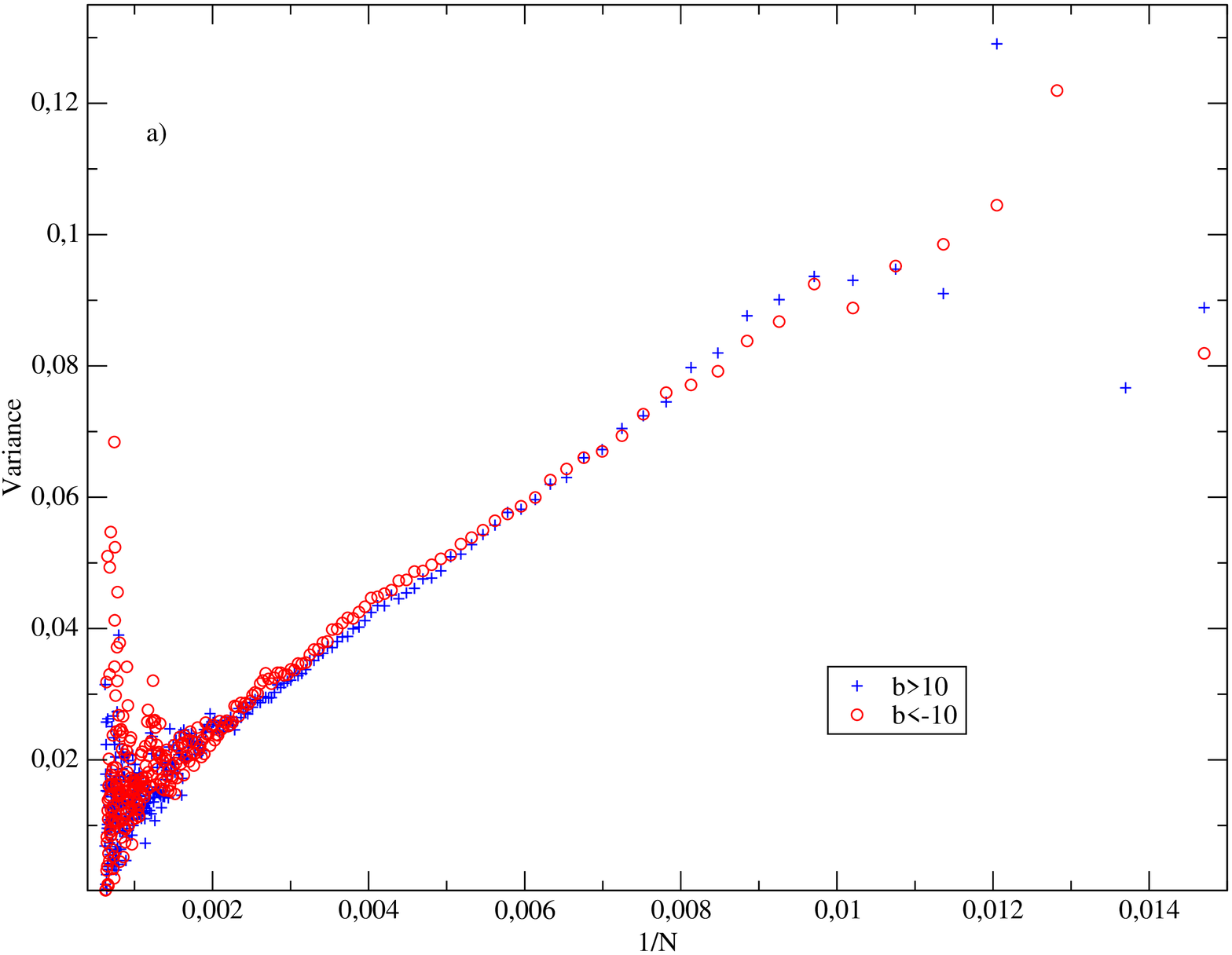}
\includegraphics[width=6.8cm,angle=270]{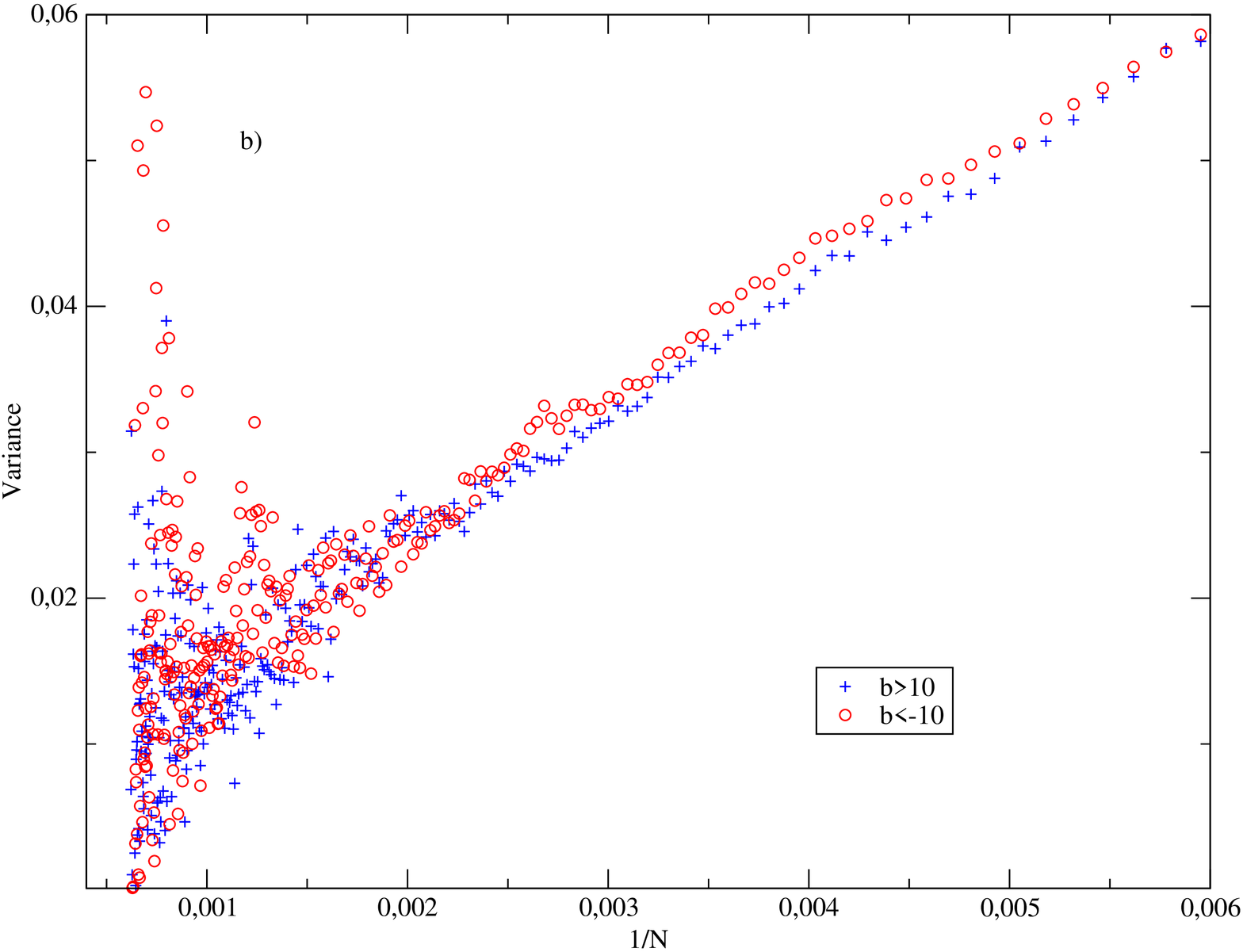}
\end{minipage}
\begin{minipage}[b]{1\textwidth}
\includegraphics[width=6.8cm,angle=270]{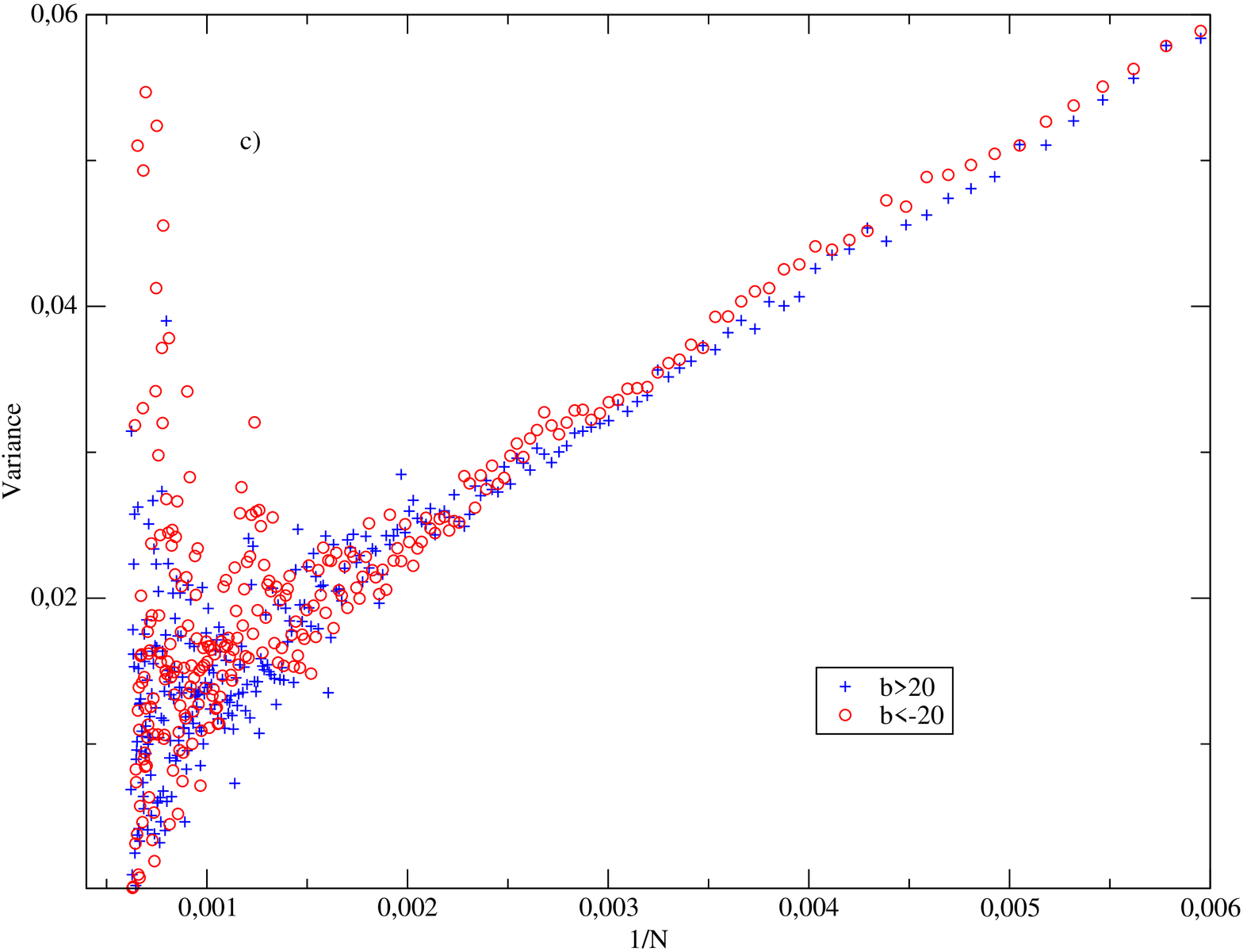}
\includegraphics[width=6.8cm,angle=270]{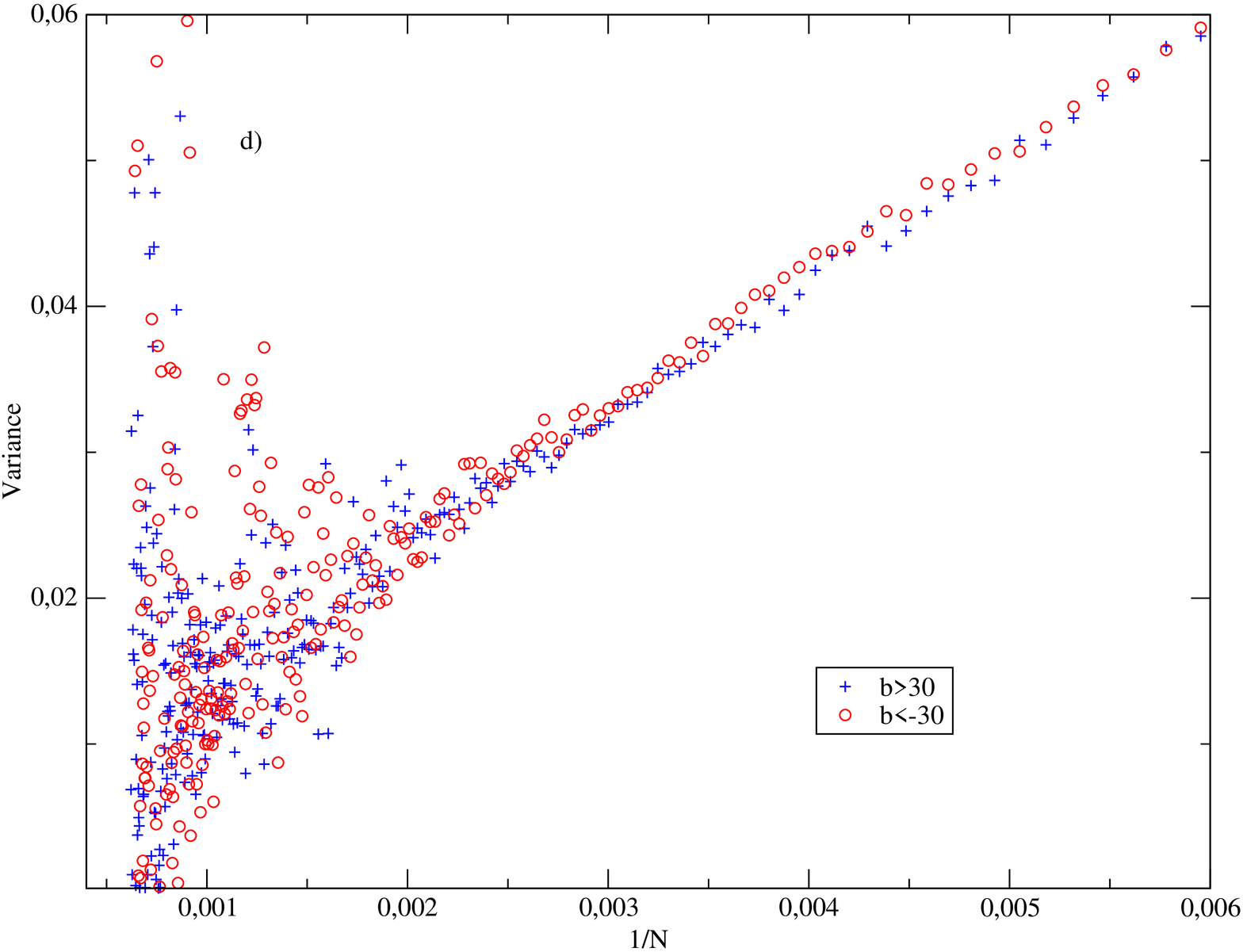}
\caption{
Values of the measured variance $\sigma^2$ versus the
number of observations $1/N$ for the V2 sky map.
}
\label{fig3}
\end{minipage}
\end{figure*}

\section{Asymmetric noise distributions of galactic hemispheres}

Apart from the above analysis of signal gradients, we studied also the 
large-scale asymmetries within the WMAP data by deriving diagrams of the measured 
signal variance across the sky versus the effective number of observations, N.  
Such diagrams, presented by Jarosik et al. (2003) for the whole sky (see their 
Fig.10),  we constructed independently for the northern and southern galactic 
hemispheres, in a couple of galactic polar-cap regions formed by removing the 
galactic equatorial strip with $|b| < 10^\circ$, $20^\circ$ and $30^\circ$ 
(Fig.~3). We used these growing galactic cuts to check if possible 
galactic foregrounds may affect the data. To derive the sample signal variance 
for a consecutive bin "$i$" centered at the number of observations $N_i$ we 
used data from all strip cells with $N$ in the respective range ($N_i-2.5$, 
$N_i+2.5$). There were also much smaller numbers of cells with very small or very 
large $N$, which can explain growing scatter at limiting parts of the plot 
presented in the upper-left panel of Fig~3.  In the other three panels in 
this figure the zoomed distributions are presented, excluding the range 
with low $N$, where the linear form of distribution is lost. In the 
figure one can easily see the south-north asymmetry in the variance distribution.  
The difference is preserved in a range of regions in the sky with varying 
number of observations. Moreover, there is a distinct tendency for decreasing 
the difference between hemispheres at higher galactic latitudes.

\begin{figure*}                                                 
\includegraphics{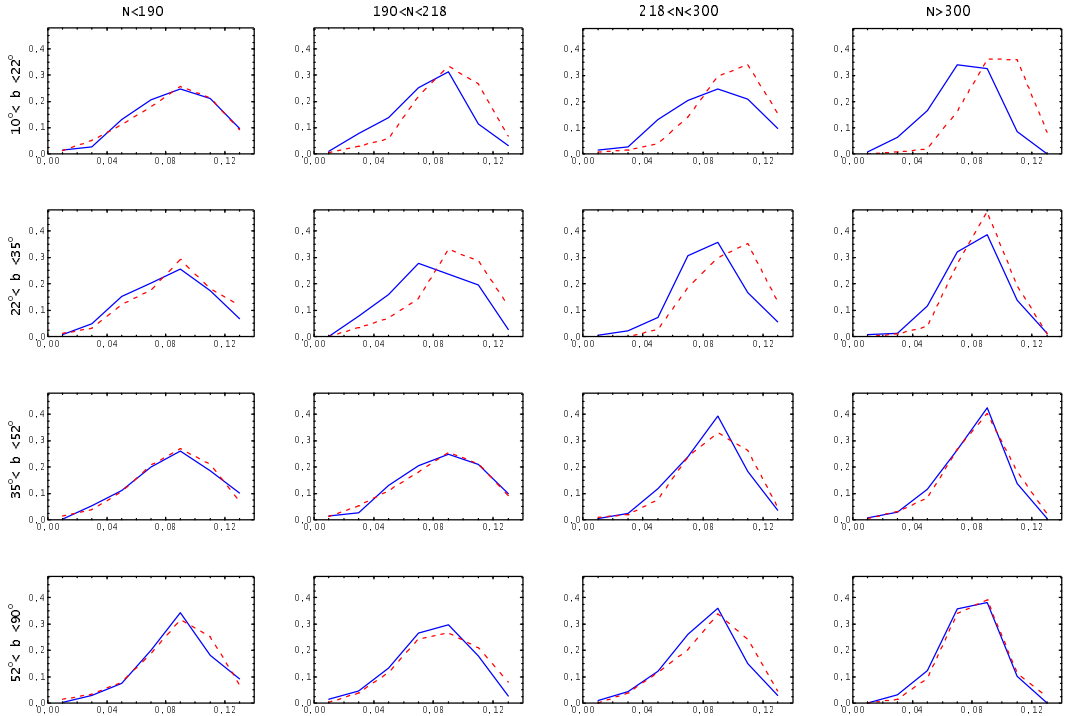}
\caption{
Histogram points connected with lines showing the signal dispersion
$\sigma_{signal} \equiv \sqrt{<\Delta T_{signal}^2>}$ distributions at 4
different galactic altitude stripes ($b \in$ [$10^\circ$, $22^\circ$] -
first row, [$22^\circ$ , $35^\circ$] - second row, [$35^\circ$, $52^\circ$]
- third row, and the polar cup [$52^\circ$, $90^\circ$] - forth row) in
different observational time intervals (N<190), 190<N<218, 218<N<300, 
N>300). The measured dispersion is presented along the horizontal
axis of each panel in 0.02 bins. In the following panels the results are
compared for the northern galactic hemisphere (full line) and for the
southern one (dashed line).
}
\label{fig4}
\end{figure*}

We performed another analysis of the signal dispersion in the channel V2. 
We considered 4 separate galactic altitude stripes ($b \in $ [$10$, $22$], [$22$, $35$], [$35$, $52$], [$52$, $90$]) and 4 ranges of the observational time ($N<190$, $190<N<218$, $218<N<300$, $N>300$), separately in both hemispheres. The individual dispersion values were derived in the following way. We assumed a strict relation to hold between dispersions of the cosmic signal (dominated by the CMB) fluctuations, 
$\sigma_{cs}$, and of the instrumental noise, $\sigma_{noise}(N) = \sigma_0/\sqrt{N}$,
constituting the measured dispersion $\sigma$ of temperature fluctuations ($\Delta T$) on the map, 

$$\sigma^2 = \sigma_{\rm cs}^2 + {\sigma_0^2 \over N} \qquad . \eqno(4.1)$$

\noindent
The noise constant for the analysed V2 channel is given by the WMAP team as 
$\sigma_0 = 2.93683$ (Bennett et al. 2003a). Our operational procedure to derive the  
individual dispersion value involves data from n ($\approx 300$) consecutive cells 
from the V2 sky map (for the given stripe and the observation number range):

$$\sigma_{\rm cs} = \sqrt{ { \left[ \sum_{i=1}^{n} (\Delta T_i)^2 \right] \over n} - {\sigma_0^2 \over N}} \qquad . \eqno(4.2)$$

\noindent
One may note that the presented algorithm, while not supposed to yield
strict statistical quantities, provides a uniform measure of fluctuation 
amplitude, allowing for comparing the sky signal in both hemispheres.  

The normalized distributions of the above dispersion values $\sigma_{\rm cs}$ 
are presented at Fig.~4. A distinct systematic trend can be observed in the 
successive panels, with the south-hemisphere distributions shifted to higher 
dispersion values. The observed difference is larger in sky stripes closer to 
the galactic plane and within these stripes is more pronounced for panels with 
larger $N$ 
(the longer observational times). However, in regions near the galactic 
poles such weak but regular shift is also observed.

\section{Summary and Conclusions}

The gradient analysis is a very efficient method in recognition of weak signal 
non-uniformities in the sky maps. By analysing the gradient maps we demonstrate 
the WMAP noise pattern resulting from non-uniformity of the observational time 
distribution. The pattern involves regions of low noise (small gradients) near 
the ecliptic poles and along the circles at the angular distance of 
$\approx 23^\circ$ from the ecliptic poles, and non-uniform extended regions 
of larger noise amplitudes (larger gradients) along the ecliptic equator. The 
pattern introduced by local variations in the observational time is much alike 
at all available frequency channels and must inevitably affect the resulting 
CMB maps, independently of the applied approach to constructing such maps from 
the original data (e.g. Bennett at al. 2003b, Eriksen et al. 2004b, Tegmark 
et al. 2003). The apparent gradient structure can "propagate'' through the 
noise filtering procedures used in attempt to 
reconstruct the CMB signal maps, 
as it is proved in section 3 by showing its imprints 
in the frequently 
presented and discussed ILC map. It suggests that analyses of 
the ILC maps, or similar maps, must be affected - besides statistical flaws 
due to the signal generation procedure and non-removed foreground effects - 
by the WMAP 
observational time pattern (cf. in this context e.g. Fig.~1 in Hansen et al. 2004c, 
Fig.~4 in Jaffe et al. 2005a and 
Jaffe et al. 2005b). One may also note that for the CMB study based on the WMAP data
(and the future Planck mission) the best regions to study are those with the 
lowest gradients around the ecliptic poles, and the lowest instrumental noise.

\begin{figure}                                                  
\includegraphics[width=8.5cm]{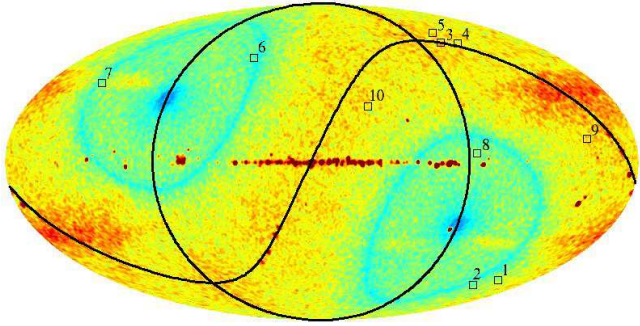}
\caption{
The V2 gradient map from Fig.~1c (but in a linear scale) with plotted characteristic
directions described in the text. For orientation, the lines are added for
the ecliptic equator and the ecliptic meridian perpendicular to the
galactic plane.
}
\label{fig5}
\end{figure}

Independently, the possible galactic foreground effects were studied by simple 
methods of analysing distributions of signal fluctuations versus the 
observational time at the WMAP V2 band. We confirm existence of the south-north 
asymmetry between hemispheres, with the larger fluctuation amplitude in the south 
hemisphere (cf. Eriksen et al. 2004a, Hansen et al. 2004a,b). We show that the 
asymmetry extends throughout a large part of the sky and diminishes at higher 
galactic altitudes. It suggests that the foreground effects may be due to some 
galactic component concentrating along the galactic plane. Furthermore, higher 
amplitude of signal fluctuations in the southern hemisphere can be, in our opinion, 
explained by the known fact that our Sun is situated approximately 20 pc north 
off this plane. Let us also note that regularity of observed effects with 
growing galactic altitude, also in the separate ranges of $N$ at Fig.~4, does not 
support the hypothesis suggesting an extragalactic origin of the asymmetry.

The findings suggest that the analysis of large-scale patterns of the 
CMB signal based on the WMAP data must be affected by observational time 
non-uniformities, which are not possible to be quite removed. Thus we expect 
that analysis results of the global CMB sky distribution can be also influenced 
by this fact, introducing the ecliptic symmetry to the data, and by the 
foreground effects, leading to some symmetry/asymmetry with respect to the 
galactic plane. Thus let us look at examples of characteristic orientations/axes 
of the universal CMB distribution discovered by the other authors (e.g. in discussion of 
non-trivial universe topology by de Oliveira-Costa et al. 2004, Weeks et al. 2004, 
Roukema et al. 2004\footnote{but Cornish et al. (2004) do not confirm non-trivial 
cosmological effects}). Such 10 directions imposed onto the gradient map in 
Fig.~5 include: 1.)  ($l=207^\circ$, $b=-59^\circ$) Vielva et al. (2004); 2) 
(222$^\circ$, -62$^\circ$) Jaffe et al. (2005a); 3) (-100$^\circ$, 60$^\circ$) 
Land et al. (2005); 4) (-110$^\circ$, 60$^\circ$) de Oliveira et al. (2004); 5) 
(252$^\circ$ , 65$^\circ$); 6) (51$^\circ$, 51$^\circ$); 7) (144$^\circ$, 38$^\circ$); 
8) (271$^\circ$, 3$^\circ$); 9) (207$^\circ$, 10$^\circ$); 10) (332$^\circ$, 25$^\circ$) 
Roukema et al. (2005). The presented directions show more or less preferential 
orientation with respect to four circles: the ecliptic equator and its meridian 
perpendicular to the galactic plane, 
plus two small circles in the distance of $23^\circ$ around the ecliptic poles, 
the sites corresponding to the local maxima of the observational time distribution. 
Also the discussed unexplained quadrupole and octopole orientations in respect to the 
ecliptic frame (cf. de Oliveira-Costa et al. 2004, Land \& Magueijo 2005, 
Jaffe et al. 2005) can be possibly explained in a natural way, if their detections 
are affected by the sky pattern due to the observational time, retained in the cleaned maps. 

In the present work we were not able to verify the supposition that several 
detected structures in the CMB maps result from the discussed instrumental or 
foreground effects, but the distinct grouping of the observed structure 
orientations within some regions of the ecliptic and galactic reference frames 
suggests that such possible effects should be properly considered.

\thanks{This work made use of the {\it WMAP} data archive and the HEALPIX
software package. We are grateful to Reinhard Schlickeiser and Andrzej Woszczyna for 
stimulating discussions. The work was performed within the research program of 
the Jagiellonian Center of Astrophysics.}

\end{document}